\documentclass[eqsecnum,showpacs,pre,twocolumn]{revtex4-1}%
\usepackage[latin1]{inputenc}
\usepackage{graphicx}
\usepackage{amsmath}
\usepackage{bm}
\usepackage{amsfonts}
\usepackage{amssymb}
\usepackage[usenames]{color}%
\begin{document}
\title{Driven surface diffusion with detailed balance and elastic
phase transitions}
\author{Hans-Karl Janssen}
\affiliation{Institut f\"ur Theoretische Physik III, Heinrich-Heine-Universit\"at, 40225
D\"usseldorf, Germany}
\author{Olaf Stenull}
\affiliation{Department of Physics and Astronomy, University of Pennsylvania, Philadelphia
PA 19104, USA}
\date{\today}

\begin{abstract}
Driven surface diffusion occurs, for example, in molecular beam epitaxy when
particles are deposited under an oblique angle. Elastic phase transitions
happen when normal modes in crystals become soft due to the vanishing of
certain elastic constants. We show that these seemingly entirely disparate
systems fall under appropriate conditions into the same universality class. We
derive the field theoretic Hamiltonian for this universality class, and we use renormalized field theory to calculate critical exponents and logarithmic corrections for several experimentally relevant quantities.
\end{abstract}
\maketitle

\section{Introduction}

Universality is the amazing phenomenon when the critical behavior arising in
different physical systems is characterized by a common set of universal
quantities. All systems belonging to a certain universality class share the
same critical exponents, universal amplitude ratios and universal scaling
functions. Typically, universality classes are determined by spatial
dimension, symmetry and the range of interaction potentials. At times, two
systems belonging to one universality class appear to be entirely disparate at
first sight. In this paper we discuss such a remarkable case, \emph{viz}. that
of driven surface diffusion under detailed balance conditions and crystalline
solids near their shear instability. We show that these respectively two and
three dimensional systems are described by a common field-theoretic
Hamiltonian. We calculate the scaling behavior of this universality class
using field-theoretic renormalization group (RG) methods.

Surface growth problems constitute an important class of generic
nonequilibrium phenomena \cite{HHZ95,Kr97,Kr02,MaMaToBa96a}. Particles are deposited on a
surface and can diffuse around on it. In
ballistic deposition processes such as molecular beam epitaxy
(MBE), desorption and bulk defect formation can often be neglected. After an
initial transient, a steady state is established which is characterized by
time-independent macroscopic properties, provided that a suitable reference
frame is chosen. If one is primarily interested in universal, large-scale,
long-time characteristics, the dynamical evolution can often be formulated in
terms of a Langevin equation which then can be analyzed using the methods of
renormalized field theory.

\begin{figure}[ptb]
\centering{\includegraphics[width=7cm]{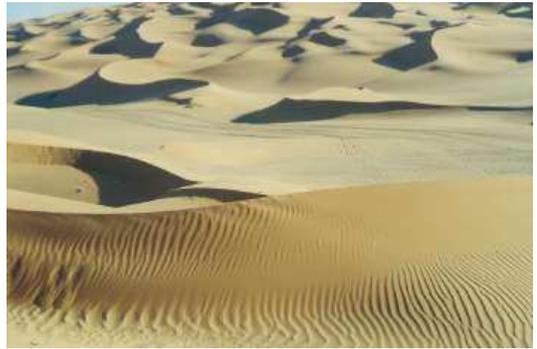}}\caption{(Color online) Sandripples in
the Lybian desert (source: Wikipedia). Steady winds drive sand particles to form longitudinal ripples on various length scales.}%
\label{fig:ripples}
\end{figure}

Several years ago, Schmittmann, Pruessner and one of us (H.K.J.)
\cite{SchmPruJa06a,SchmPruJa06b}, denoted in the following as SPJ, extended a
model by Marsili, Maritan, Toigo, and Banavar (MMTB) \cite{MaMaToBa96b}, to
describe ballistic deposition processes with oblique particle incidence. In broad terms, their approach can be described as follows: It is assumed that there is no shadowing, i.e., that there are no overhangs so that the growing surface can be described in terms of a height variable $h(\mathbf{r},t)$ (so-called Monge representation). Furthermore, it is assumed that particle desorption and bulk defect
formation can be neglected, so that the (deterministic) surface relaxation processes
are particle-number conserving. Then, the time rate of change of the surface-hight is formulated
in an idealized continuum description~\cite{WoVi90,Vi91,LaDSa91,Ja96,EsKo12,ShePru12} in terms of the
divergence of a current. Since the absolute hight of the surface is irrelevant, these currents are constructed from the slope of the height $h$ and its
derivatives. Next, shot noise modeling the random deposition of the particles is
incorporated into the growth equation by adding a non-conserving stochastic term which dominates the conserving diffusional noise. Since the oblique particle beam selects a preferred (\textquotedblleft
longitudinal\textquotedblright) direction in the substrate plane, the
resulting Langevin equation is necessarily anisotropic, and the primary
relaxation mechanism is driven surface diffusion. The interplay of interatomic
interactions and kinetic effects, such as Ehrlich-Schwoebel barriers,
generates an anisotropic effective surface tension which can become very small
or even negative. Due to the anisotropy, these mechanisms lead to a phase
space with four different regimes, and the potentially scale-invariant
behaviors in these regimes were studied by SPJ . The most interesting case
turned out to have a transversal instability which leads to the generation of
longitudinal ripples via a
continuous phase transition in contrast to a longitudinal instability which leads to the generation of transversal ripples
by a (fluctuation induced) first order transition. Such ripple structures are commonly found in sandy deserts where they are
generated under the influence of a steady wind as the driving force, see Fig.~\ref{fig:ripples}. It is worth noting that MMTB and SPJ formulated their model under the
assumption of an invariance with respect to tilts of the surface. However, in
a real setup, the orientation of the basic surface relative to the incident
particle beam is generally fixed, and hence it is useful to generalize their
model to go beyond the limiting tilt-invariant case. Here, in the present
work, we do \emph{not} assume tilt invariance. We will see as we move along
that the resulting growth equation is in detailed balance
when an appropriate co-moving coordinate system is used and a certain model parameter vanish. Note that we use 
the term \textquotedblleft detailed balance\textquotedblright\ not in a microscopic sense \cite{Ja92}. Even if the
microscopic laws lack detailed balance, coarse graining and the following neglect of irrelevant terms can lead to detailed balance in the asymptotic semi-macroscopic equation of motion for a driven system~\cite{JaSchm86,SchmZi95}.
Under the assumption that this is the case, we derive a quasi-static Hamiltonian $\mathcal{H}$ which describes the
stationary distribution, and whose basic ingredients are derivatives of the surface height field $h(\mathbf{r})$.

The elastic free energy of crystalline solids with hexagonal or tetragonal
(including cubic) symmetry shows an instability against shear when the elastic
constant $C_{44}$ goes to zero. The crystallographic unit cell undergoes an
elastic deformation with a certain linear combination of components of the
strain tensor as an order parameter. In the case that no third order
invariants of the order parameter are possible
(or vanish by chance in the cubic case), the transition to
the structurally distorted crystal is continuous, and the order parameter can
be reduced to the spatial derivatives of the displacement field $u$ along the
crystallographic symmetry axis \cite{Cow76,FoIrSchw76a,FoIrSchw76b,SchwTa96}.
Augmenting the stretching energy with stabilizing bending terms and then
reducing it to its parts that are relevant in the RG
sense, we set up a generalized elastic energy that is of the same form as the
aforementioned quasi-static Hamiltonian $\mathcal{H}$ with the displacement
field $u(z,\mathbf{r}_{\bot})$ taking on the role of the height field
$h(z,\mathbf{r}_{\bot})$. Note, however, there is one important distinction
between the two systems. Whereas in the driven diffusion problem, both the
longitudinal component $z$ and the transversal components $\mathbf{r}_{\bot}$
lie in the plane of the substrate, i.e., in the plane orthogonal to the height
$h$, the coordinate $z$ in the elastic problem lies in the direction of the
crystal axis and is therefore parallel to the displacement $u$. Hence, the
physical dimension of $\mathbf{r}_{\bot}$ is $1$ in driven surface diffusion
and $2$ in the elastic transition.

The outline of the present paper is as follows: In Sec.~\ref{sec:model} we set
up field theoretic models for driven surface diffusion with detailed balance
and elastic phase transitions, respectively, through augmenting and modifying
the existing models for these systems. We unify the two field theoretic models
by formulating a Hamiltonian that commonly describes both systems. In
Sec.~\ref{sec:RPT} we present our renormalized perturbation theory. We sketch
our diagrammatic calculation, discuss the resulting renormalization group and
its flow. In Sec.~\ref{sec:scalingAndExponents}, we derive the scaling forms for several experimentally measurable quantities. For driven surface diffusion, we calculate critical exponents, and for elastic transitions we calculate logarithmic corrections to the leading mean-field behavior. In Sec~\ref{sec:concludingRemarks} we give concluding remarks.
Appendix~\ref{app:calcDiagrams1Loop} contains some details on our calculation
of Feynman diagrams. Appendix~\ref{app:addRen} discusses some intricacies related to the fact that the calculation of the specific heat for the elastic transitions requires additive renormalizations.

\section{The Model}

\label{sec:model}

\subsection{Driven Surface Diffusion}

Here, we set up our model for driven surface diffusion following the work of
MMTB~\cite{MaMaToBa96a,MaMaToBa96b} and SPJ. As a starting
point, let us review the Langevin equation presented by SPJ. In the
\^{I}to-interpretation, this Langevin equation describes the hight
$h(\mathbf{r},t)$ of a $d$-dimensional growing hyper-surface as
\begin{equation}
\partial_{t}h=J_{0}+\mathbf{J}\cdot\nabla h-\nabla\cdot\mathbf{j}+\zeta\,.
\label{GenLang}%
\end{equation}
The incident particle beam has components $J_{0}$ and $\mathbf{J}$ normal and parallel
to the substrate plane, respectively. As discussed in detail by MMTB and SPJ, the oblique particle incidence induces a fundamental anisotropy into the system which breaks the full rotational symmetry within the $d$-dimensional space of the substrate. To account for this anisotropy, it is useful to choose a coordinate system in which a particular axis, say the $z$-axis, is aligned with $\mathbf{J}$ and then to write the spatial coordinates as 
$\mathbf{r}=(z,\mathbf{r}_{\bot})$.  The two uniform growth terms, the first two terms on the right hand side of  Eq.~(\ref{GenLang}), can be removed by a Galilei transformation $h(\mathbf{r},t)\mapsto h(\mathbf{r+J}t,t)+J_{0}t$. In the following, we will work in this co-moving frame. $\mathbf{j} = (j_\parallel, \mathbf{j}_\perp)$ is the surface current, and we model its longitudinal and transversal components following SPJ by
\begin{align}
j_{\parallel}  &  =\lambda\Big\{\partial_{\parallel}\bigl(-\tau_{\parallel
}h+\kappa_{\parallel}\partial_{\parallel}^{2}h+\kappa_{\times}\nabla_{\perp
}^{2}h\bigr)\nonumber\\
&  \qquad\qquad+\frac{g_{\parallel}}{2}\bigl(\partial_{\parallel}%
h\bigr)^{2}+\frac{g_{\perp}}{2}\bigl(\nabla_{\perp}h\bigr)^{2}%
\Big\}\,,\label{parcurr}\\
\mathbf{j}_{\perp}  &  =\lambda\Big\{\nabla_{\perp}\bigl(-\tau_{\perp}%
h+\kappa_{\perp}\nabla_{\perp}^{2}h+\kappa_{\times}\partial_{\parallel}%
^{2}h\bigr)\nonumber\\
&  \qquad\qquad+g_{\times}\bigl(\nabla_{\perp}h\bigr)\,\bigl(\partial
_{\parallel}h\bigr)\Big\}\,.
 \label{perpcurr}%
\end{align}
Note that under the assumptions made by MMTB and SPJ, $g_{\times}=-g_{\perp}$. As we will discuss in more detail below, we will work under different assumptions here. The Langevin noise $\zeta$ accounts for the randomness in the particle deposition. Because this shot noise adds particles to the surface, it is commonly modeled as non-conserving noise with correlations of the form
\begin{equation}
\overline{\zeta(\mathbf{r},t)\zeta(\mathbf{r}^{\prime},t^{\prime})}%
=2\lambda\,\delta(\mathbf{r}-\mathbf{r}^{\prime})\delta(t-t^{\prime})\,.
\label{shot-noise}%
\end{equation}
In principle, conserved diffusional noise could also be incorporated into the model. However, it turns out to be irrelevant compared to the non-conserved noise. 

The character of the surface produced by the process depends on the values of the critical parameters $\tau_{\Vert}$ and $\tau_{\bot}$. Our model leads to qualitatively the same phase diagram in the space spanned by $\tau_{\Vert}$ and $\tau_{\bot}$ as the SPJ model, see Fig.~1 of Ref.~\cite{SchmPruJa06b}. This phase diagram contains, in particular, a continuous transition from isotropic, Edwards-Wilkinson~\cite{EW1982} type behavior to a state characterized by longitudinal ripples for $\tau_{\bot}\rightarrow0$ and $\tau_{\Vert}>0$. In the following, we will focus on this transition. The other transitions in the phase diagram which includes a fluctuation induced discontinuous transition to a surface with transversal ripples for $\tau_{\Vert}\rightarrow0$ and
$\tau_{\bot}>0$ and a multi critical point for $\tau_{\Vert}\rightarrow0$ and $\tau_{\bot} \rightarrow0$ are set aside for future work.

Now, we review the symmetries and the physical contents of the Langevin equation~(\ref{GenLang}) with its ingredients (\ref{parcurr}) to (\ref{shot-noise}) in some more depths. The work by MMTP and SPJ assumed that the physics of the system is invariant under an infinitesimal tilt of the surface by an angle $\mathbf{\omega}$, $h(z,\mathbf{r}_{\bot})\mapsto h(z,\mathbf{r}_{\bot})+\mathbf{\omega\cdot r}_{\bot}$. The assumption of this invariance enforces the relation $g_{\times}=-g_{\perp}$ mentioned above~\cite{footnote1}.
However, this tilt-invariance is non-physical since the orientation of the substrate plane
is fixed by the normal component $J_{0}$ of the incident current. In the
following, we will refrain from imposing this condition. Stability demands
positive $\kappa_{\Vert}$ and $\kappa_{\bot}$ as well as $\kappa_{\times}%
\geq-2\sqrt{\kappa_{\Vert}\kappa_{\bot}}$. At the transition for $\tau_{\bot}\rightarrow0$
and $\tau_{\Vert}>0$ that we are interested in,  $\kappa_{\Vert}$ and $\kappa_{\times}$ are irrelevant.
The positive parameters $\kappa_{\bot}$ and $\tau_{\Vert}$ can be set equal to
$1$ and a dimensionless positive parameter $\rho^{2}$, respectively, via a
simple rescaling of the coordinates $\mathbf{r}=(z,\mathbf{r}_{\bot})$ and
$\tau_{\bot}=\tau$. Note that after this transformation, the coordinates, the time and the "temperature" scale as $\mathbf{r}_{\bot}\sim\mu^{-1}$, $z\sim\mu^{-2}$, $t\sim(\lambda\mu^{4})^{-1}$ and $\tau\sim\mu^{2}$, respectively, where $\mu$ is an inverse external length scale
appropriate to discuss the infrared limit that will be used in the renormalization
program to follow. It thus follows from the Langevin equations,
Eqs.~(\ref{GenLang}) and (\ref{shot-noise}), that the shot noise $\zeta$ and
the height $h$ scale as $\zeta\sim\lambda\mu^{(d+5)/2}$ and $h\sim
\mu^{(d-3)/2}$. Furthermore, the coupling constants scale as $g_{\parallel
}\sim\mu^{-(d+1)/2}$ and $g_{\perp}\sim g_{\times}\sim\mu^{(3-d)/2}$ which
shows that $g_{\parallel}$ is irrelevant and that the upper critical dimension is
$d_{c}=3$. 

Returning to the tilt-invariance and the fact that we do not assume it in the present work, we note that there is a possible contribution to $\mathbf{j}_{\perp}$ of the form $f(\nabla_{\bot}^{2}h)(\nabla_{\bot}h)/6$ that was omitted by MMTB and SPJ on grounds of tilt-invariance, and we have to consider it here. Power counting reveals that $f\sim\mu^{3-d}$, i.e., this term is relevant. Even if we omitted it at the onset, it would be generated by the RG. Here, we include it in our model at this stage.

We are interested here in driven surface diffusion with detailed balance so that the statistical properties of the steady state can be derived from a \textquotedblleft free energy\textquotedblright or Hamiltonian $\mathcal{H}_{\text{surf}}$. For this to hold true, we have to be able to express the components of the surface current through functional derivatives of $\mathcal{H}_{\text{surf}}$. Because the absolute height of the surface should not matter, i.e., $h\mapsto h+h_{0}$ with a constant $h_{0}$ should be a symmetry, the functional variation of $\mathcal{H}_{\text{surf}}$ with respect to $h$ should be of the form
\begin{align}
\delta \mathcal{H}_{\text{surf}}&= \int d^{d}x \, \frac{\delta\mathcal{H}_{\text{surf}}}{\delta
 h} \, \delta h = \int d^{d}x \, \mathbf{j} \cdot \nabla \delta h 
 \nonumber \\
 & = \int d^{d}x  \left(  j_{\parallel} \partial_\parallel \delta h +  \mathbf{j}_\perp \cdot \nabla_\perp  \delta h  \right)
 \,, \label{variationHamsurf}%
\end{align}
which defines the surface current  $\mathbf{j} = (j_\parallel, \mathbf{j}_\perp)$ up to divergence-free  additive part. It is easy to see that this is possible if we equalize the coupling constants
$g_{\times}=g_{\perp}=g$, and in the following we will do so. 
Then the Hamiltonian is given by%
\begin{align}
\mathcal{H}_{\text{surf}}  &  =\int d^{d}x\,\Big\{\frac{\rho^{2}}{2}%
(\partial_{\Vert}h)^{2}+\frac{\tau}{2}(\nabla_{\bot}h)^{2}+\frac{1}{2}%
(\nabla_{\bot}^{2}h)^{2}\nonumber\\
&  +\frac{g}{6}(\partial_{\Vert}h)(\nabla_{\bot}h)^{2}+\frac{f}{24}%
\bigl((\nabla_{\bot}h)^{2}\bigr)^{2}\Big\}\,. \label{Hamsurf}%
\end{align}
For further details and background on detailed balance in the context of driven surface diffusion, we refer to Ref.~\cite{Kr97}.

There is another symmetry present that we have not mentioned yet, viz.\ the so-called up-down symmetry. This symmetry, which physically means that mounds and valleys have the same form, translates for MBE with oblique incidence to a form-invariance under the simultaneous inversion $h\mapsto-h$ and $z\mapsto-z$. This type of up-down symmetry was assumed by MMTB and SPJ, and we also assume it here. However, even though it is frequently assumed in MBE theory, up-down symmetry has to be taken with a grain of salt because the particle beam sets a preferred direction in space. Therefore, the up-down symmetry is strictly speaking more appropriate for interface than for surface models. We keep this symmetry here firstly because we are primarily interested in the connection between and the mutual properties of driven surface diffusion and elastic phase transitions and secondly because the up-down symmetry simplifies the model. In a sense, this simplified model can be viewed as a leading order approximation for MBE with oblique incidence that can be improved later by including terms that break the up-down symmetry. In fact, we plan to study the effects of breaking the up-down symmetry in the future. Note that this symmetry has been discussed in the literature~\cite{Kr02} as a prerequisite for having a \textquotedblleft free energy\textquotedblright  in this kind of system. However, we can add a further coupling proportional to $(\nabla_{\bot}^{2}h)(\nabla_{\bot}h)^{2}$ to the Hamiltonian
(\ref{Hamsurf}) which explicitly destroys up-down symmetry. Only in the
one-dimensional case this coupling is a total gradient and therefore vanishes
after integration over the full one-dimensional space. 

The Langevin equation (\ref{GenLang}) now takes the form
\begin{equation}
\partial_{t}h=-\lambda\frac{\delta\mathcal{H}_{\text{surf}}}{\delta h}+\zeta\,
\label{SpecLang}%
\end{equation}
in the comoving-frame introduced above. To generate correlation and response
functions of this stochastic process, it lends itself to reformulate the
Langevin equation as a dynamical response functional $\mathcal{J}%
_{\text{surf}}$ so that averages can be derived from path integrals with the
statistical weight $\exp(-\mathcal{J}_{\text{surf}})$. Via introducing a
response field $\tilde{h}(\mathbf{r},t)$, the functional $\mathcal{J}%
_{\text{surf}}$ (in \^{I}to-discretization) that we extract with standard methods~\cite{Ja76,DeDo76,Ja92} from
Eqs.~(\ref{SpecLang}) and (\ref{shot-noise}) reads
\begin{equation}
\mathcal{J}_{\text{surf}}[\tilde{h},h]=\int d^{d}x\int dt\ \lambda
\Big\{\tilde{h}\Big(\lambda^{-1}\partial_{t}h+\frac{\delta\mathcal{H}%
_{\text{surf}}[h]}{\delta h}\Big)-\tilde{h}^{2}\Big\}\,, \label{def_J}%
\end{equation}
and shows that our driven surface diffusion model obeys detailed balance
\cite{Ja92}. The fluctuation-dissipation theorem (FDT) holds, and all
stationary expectation values can be calculated from path-integrals with the
statistical weight $\exp(-\mathcal{H}_{\text{surf}})$.

 It is a well established fact that in all dynamical field theories that are invariant under
a field shift of the type $h\mapsto h+h_{0}$ with a constant $h_{0}$, the
dynamical exponent is given by $z=4-\eta$, where $\eta/2$ is the anomalous
scaling exponent of the order parameter field $h$. As it should, the same applies here. This can be verified in detail by noting that in the full dynamical field theory based on the response functional~(\ref{def_J}) one does not need to introduce renormalization factors for the $\tilde{h}\partial_{t}h$ and $\lambda\tilde{h}^{2}$ terms when using minimal renormalization, i.e., dimensional regularization in conjunction with minimal subtraction. Hence, we can calculate all of the critical exponents of driven surface diffusion with detailed balance from $\mathcal{H}_{\text{surf}}$ including the dynamical exponent.

\subsection{Elastic Phase Transitions}

An elastic phase transitions is a structural phase transition in which the
crystallographic unit cell undergoes an elastic deformation when a normal mode
becomes soft in response to the vanishing of an elastic constant or
combination of elastic constants. In this section, we essentially follow the
early work of Cowley \cite{Cow76}, and Folk, Iro, and Schwabl
(FIS)~\cite{FoIrSchw76a,FoIrSchw76b}. For background information on elastic
phase transitions, we refer to these original papers as well as reviews by
Bruce and Cowley~\cite{BruceCowley81}, Rao and Rao~\cite{RaoRao87}, and
Schwabl and T\"{a}uber \cite{SchwTa96}.

In general, the order parameter for such a transition is a certain component
of the Lagrange strain tensor or a linear combination of its components.
Depending on the symmetry of a crystal, its Lagrange elastic energy density
can be unpleasantly complicated even at harmonic order due to the presence of
a host of different elastic constants. Here, we restrict ourselves to crystals with tetragonal or
higher symmetry, i.e., crystals whose harmonic Lagrange energy density can be
obtained from the tetragonal case by imposing relations between certain
elastic constants. As we will see below, crystals with lower symmetry such as orthorhombic or monoclinic symmetry have simple mean field behavior at their respective elastic transition and are hence less interesting from the standpoint of critical phenomena than the higher-symmetry crystals we focus on here.
When generalized to $d$ spatial dimensions, the Lagrange energy density of
tetragonal solids to second order in the strains reads
\begin{align}
f_{\text{el}}  &  =\frac{C_{11}}{2}\sum_{\alpha}u_{\alpha\alpha}^{2}%
+\frac{C_{33}}{2}u_{dd}^{2}+2C_{44}\sum_{\alpha}u_{\alpha d}^{2}\nonumber\\
&  +2C_{66}\sum_{\alpha<\beta}u_{\alpha\beta}^{2} +C_{12}\sum_{\alpha<\beta
}u_{\alpha\alpha}u_{\beta\beta}+C_{13}\sum_{\alpha}u_{\alpha\alpha}u_{dd}\,.
\label{frEn}%
\end{align}
where $\alpha,\beta=1,\ldots,d-1$, and $d=3$ for a physical solid. The
Lagrange strain tensor (see, e.g., Refs.~\cite{Landau-elas,tomsBook}) is defined via the derivatives of
the displacement field $u_{\mu}$ with respect to the reference space
coordinates of the undeformed solid,
\begin{equation}
u_{\mu\nu}=\frac{1}{2}\bigl(\partial_{\mu}u_{\nu}+\partial_{\nu}u_{\mu
}+\partial_{\mu}u_{\lambda}\partial_{\nu}u_{\lambda}\bigr)\,, \label{VerzTens}%
\end{equation}
with $\mu,\nu=1,\ldots,d$. If we specialize the elastic constants $C_{ij}$ to
$C_{33}=C_{11}$, $C_{66}=C_{44}$, $C_{13}=C_{12}$ or $C_{66}=(C_{11}%
-C_{12})/2$, Eq.~(\ref{frEn}) reduces to the harmonic energy density of a
cubic or hexagonal crystal, respectively. If all these conditions hold,
Eq.~(\ref{frEn}) describes an isotropic solid. Moreover, the hexagonal case
with the additional relation $C_{11}C_{33}=C_{13}^{2}$ reduces to the
isotropic case after a suitable rescaling of the coordinates.

Because of crystal anisotropy, sound velocities depend on the direction of
propagation, and modes become soft only in certain selected directions,
so-called soft sectors, which may be one- or two-dimensional in 3$d$ crystals.
The type of the soft modes depends on which elastic constant or combination of
elastic constants vanishes. In the following, we focus on the shear
instability that occurs when $C_{44}$ goes to zero. The accompanying elastic
phase transition is continuous if there are no third order couplings of the
corresponding order parameter. We assume that this is the case. The upper
critical dimension of the phase transition is $d_{c}=2+m/2$, where $m$ is the
dimension of the soft mode sector in Fourier space, as was shown by FIS. This
implies that, from the standpoint of critical phenomena, $m=d-1$ is the most
interesting case since otherwise the upper critical dimension is lower then
the physical dimension $3$, and the transition is simply be described by
mean-field-theory. In the remainder, we focus on $m=d-1$. In this case, the
$(d-1)$-dimensional soft sector is orthogonal to the polarization of the
corresponding critical sound waves, i.e., to the direction of the
symmetry-axis in Fourier space. Following FIS, we radically reduce the
displacement field to the critical displacement alone%
\begin{equation}
u_{\mu}=u\delta_{\mu,d}\,. \label{Reduz}%
\end{equation}
This step is justified if there are no third order couplings in the critical strain. Such couplings can be ruled out on symmetry grounds for the crystals we focus on except for the cubic case. For cubic crystals, we assume that third order couplings are absent for some given reason, e.g., chance. Otherwise, the elastic transition becomes discontinuous. We will comment on this issue and possible couplings to the non-critical displacements $u_{\alpha}$ further
below.

To figure out which parts of the elastic energy are relevant in the sense of
the RG, it is useful to rank the different terms according to the order at
which they contribute in the long-wavelength limit. To this end, let us
introduce the smallness parameter $\ell$, and set the derivatives to%
\begin{equation}
\partial_{\alpha}\rightarrow\ell\partial_{\alpha}\,,\qquad\partial
_{d}\rightarrow\ell^{2}\partial_{d}\,. \label{Klein}%
\end{equation}
With the reduction (\ref{Reduz}), the strain tensor becomes%
\begin{align}
2u_{\alpha\beta}  &  \rightarrow\ell^{2}\partial_{\alpha}u\partial_{\beta
}u\,,\nonumber\\
2u_{\alpha d}  &  \rightarrow\ell\partial_{\alpha}u+\ell^{3}\partial_{\alpha
}u\partial_{d}u\,,\nonumber\\
2u_{dd}  &  \rightarrow2\ell^{2}\partial_{d}u+\ell^{4}\bigl(\partial
_{d}u\bigr)^{2}\,, \label{SkalStrain}%
\end{align}
and we obtain the free energy density (\ref{frEn})%
\begin{align}
f_{\text{el}}  &  \rightarrow\ell^{2}\frac{C_{44}}{2}\bigl(\nabla_{\bot
}u\bigr)^{2}+\ell^{4}\Big\{\frac{C_{33}}{2}\bigl(\partial_{\Vert}%
u\bigr)^{2}\nonumber\\
&  +\frac{C_{13}+2C_{44}}{2}\bigl(\partial_{\Vert}u\bigr)\bigl(\nabla_{\bot
}u\bigr)^{2}+\frac{C_{12}+2C_{66}}{8}\bigl(\bigl(\nabla_{\bot}u\bigr)^{2}%
\bigr)^{2}\nonumber\\
&  +\frac{C_{11}-C_{12}-2C_{66}}{8}\sum_{\alpha}\bigl(\partial_{\alpha
}u\bigr)^{4}\Big\}\nonumber\\
&  +\ell^{6}\Big\{\frac{C_{33}}{2}\bigl(\partial_{\Vert}u\bigr)^{3}%
+\frac{C_{13}+2C_{44}}{4}\bigl(\partial_{\Vert}u\bigr)^{2}\bigl(\nabla_{\bot
}u\bigr)^{2}\Big\}\nonumber\\
&  +\ell^{8}\frac{C_{33}}{8}\bigl(\partial_{\Vert}u\bigr)^{4}\,,
\label{SkalFreeEn}%
\end{align}
where we have set $\partial_{\Vert} = \partial_d$ and $\nabla_{\bot} = (\partial_{\alpha})$.
Now we see that it is appropriate to rescale the vanishing elastic constant
$C_{44}$ by%
\begin{equation}
C_{44}\rightarrow\ell^{2}C_{44}\,, \label{SkalC44}%
\end{equation}
and reduce $f_{el}$ to the leading terms proportional to $\ell^{4}$:%
\begin{align}
f_{\text{el},\text{red}}  &  =\frac{C_{44}}{2}\bigl(\nabla_{\bot}%
u\bigr)^{2}+\frac{C_{33}}{2}\bigl(\partial_{\Vert}u\bigr)^{2}+\frac{C_{13}}%
{2}\bigl(\partial_{\Vert}u\bigr)\bigl(\nabla_{\bot}u\bigr)^{2}\nonumber\\
&  +\frac{C_{12}+2C_{66}}{8}\bigl(\bigl(\nabla_{\bot}u\bigr)^{2}%
\bigr)^{2}\nonumber\\
&  +\frac{C_{11}-C_{12}-2C_{66}}{8}\sum_{\alpha}\bigl(\partial_{\alpha
}u\bigr)^{4}\,. \label{RedElaEn}%
\end{align}
Of course, the anisotropic last term in $f_{el,red}$ with hypercubic symmetry
vanishes in the hexagonal case.

At this stage, a comment regarding higher than harmonic terms in the initial Lagrange energy density~(\ref{frEn}) is warranted. As long as we ignore the uncritical displacement $u_{\alpha }$, the addition to the right hand side of Eq.~(\ref{frEn}) of third and fourth order terms in the strains leads to exactly the same form of $f_{\text{el},\text{red}}$ as given in Eq.~(\ref{RedElaEn}). The only effect of the higher order strain terms is to modify the constants decorating the individual terms in Eq.~(\ref{RedElaEn}), and hence their omission is justified. However, if we included the noncritical displacements which scale as $u_{\alpha}\rightarrow\ell u_{\alpha }$, then already the harmonic strain terms in the Lagrange elastic energy density would lead to couplings of the form $\ell^{4}\partial_{\alpha }u_{\beta}\partial_{\alpha}u_{\beta}$,  $\ell^{4}\partial_{\alpha}u_{\beta
}\partial_{\beta}u_{\alpha}$ and $\ell^{4}\partial_{\alpha}u_{\beta}\partial_{\alpha }u\partial_{\beta}u$ which contribute for $\ell\rightarrow0$ at the same order in $\ell$ as the terms retained in Eq.~(\ref{RedElaEn}).

Thus far, we only included stretching in our discussion of elastic energy
densities. However, since we are interested in the limit $C_{44} \to0$ where
there is an elastic instability, bending terms will be important as they
stabilize the crystal. The bending terms that are relevant here are
\begin{equation}
f_{\text{bend}, \text{red}}=\frac{K_{1}}{2}\bigl(\nabla_{\bot}^{2}%
u\bigr)^{2}+\frac{K_{2}}{2}\sum_{\alpha}\bigl(\partial_{\alpha}^{2}%
u\bigr)^{2}\,, \label{RedBendEn}%
\end{equation}
with $K_{2}=0$ in the hexagonal case. Adding stretching and bending
contributions, we obtain the model elastic energy or Hamiltonian 
\begin{equation}
\mathcal{H}_{\text{el}}=\int d^{d}x\,\Big\{f_{\text{el}, \text{red}%
}+f_{\text{bend}, \text{red}}\Big\}\,, \label{Hel}%
\end{equation}
which is, up to the anisotropic cubic terms which are absent in the hexagonal
case, of the same form as the Hamiltonian (\ref{Hamsurf}) for driven surface
diffusion. This is the Hamiltonian that we will study, after suitable
rescalings and renamings, in our renormalized perturbation theory.

Form this point on, we will disregard the hyper-cubic terms, i.e., the last term in the stretching energy density~(\ref{RedElaEn}) and the term proportional to $K_2$ in the bending energy density~(\ref{RedBendEn}). As mentioned above, crystals with cubic symmetry are expected to have anomalous scaling behavior and are hence certainly interesting, however, the presence of the hyper-cubic terms makes the diagrammatic perturbation calculation significantly more difficult. Though we have done this calculation, we feel that a useful discussion of the cubic case goes beyond the scope of the present paper. We will cover elastic transitions in cubic crystals in a future paper~\cite{JanSteUnpub}.

Before we move on, let us make some comments on the relation of our Hamiltonian $\mathcal{H}_{\text{el}}$ to other, prominent model elastic energies that have been discussed in the literature. First, for $C_{13} = 0$ (and $K_2 = 0$), $\mathcal{H}_{\text{el}}$ reduces to one of the two models studied by FIS (their model II). We will refer to this model as the FIS model. Secondly, in the
limiting case of a fully isotropic crystal with $C_{12}=C_{13}$,
$C_{44}=C_{66}$, $C_{11}=C_{33}=C_{12}+2C_{44}$, we have%
\begin{equation}
f_{\text{el},\text{red}}^{(\text{iso})}=\frac{C_{11}}{2}I[u]^{2}\,,
\label{isotr}%
\end{equation}
with%
\begin{equation}
I[u]=\Big\{\frac{C_{44}}{C_{11}}+\bigl(\partial_{\Vert}u\bigr)+\frac{1}%
{2}\bigl(\nabla_{\bot}u\bigr)^{2}\Big\}\,, \label{Invar}%
\end{equation}
up to a total derivative. $I[u]$ is invariant under a type of a Galilean
transformation,
\begin{equation}
u(\mathbf{r)\mapsto}u(\mathbf{r-v}z\mathbf{)+v\cdot r\,,} \label{infGal-Trafo}%
\end{equation}
with infinitesimal transversal $\mathbf{v}$. When augmented with bending
the bending term proportional to $K_1$, this isotropic elastic energy density leads to a Hamiltonian
$\mathcal{H}_{el}$ that is equivalent to that considered by Grinstein und
Pelcovits (GP) in their seminal work on the anomalous elasticity of smectic
liquid crystals~\cite{GriPel81}. We will refer to this model as the GP model. Jointly, we will refer to the FIS and GP models as the special models.

\subsection{Field theoretic Hamiltonian}

To further emphasize that both driven surface diffusion with detailed balance
and elastic transition are described by the same field theoretic Hamiltonian,
we rescale the coordinates and $u$ in Eq.~(\ref{Hel}). After renaming the
order parameter field and the coupling constants in the rescaled version of
Eq.~(\ref{Hel}) as well as renaming the order parameter field $h$ in Eq.~(\ref{Hamsurf}),
both our model systems are described by the Hamiltonian
\begin{align}
\mathcal{H}  &  =\int d^{d}x\,\Big\{\frac{\rho^{2}}{2}(\partial_{\Vert}%
s)^{2}+\frac{\tau}{2}(\nabla_{\bot}s)^{2}+\frac{1}{2}(\nabla_{\bot}^{2}%
s)^{2}\nonumber\\
&  +\frac{g}{6}(\partial_{\Vert}s)(\nabla_{\bot}s)^{2}+\frac{f}{24}%
\bigl((\nabla_{\bot}s)^{2}\bigr)^{2}\Big\}\,. \label{Ham}%
\end{align}
where $s(\mathbf{r})$ is the common order parameter field.

Next, let us discuss the symmetry transformations of this Hamiltonian and the
combinations of its parameters that are invariant under the respective
transformation. Knowing these combination will come in handy later on when we
analyze the RG flow. This Hamiltonian is invariant under the longitudinal
scale transformation
\begin{align}
z  &  \mapsto\alpha^{2}z\,,\qquad s\mapsto\alpha^{-1}s\,,\nonumber\\
\rho &  \mapsto\alpha^{2}\rho\,,\qquad g\mapsto\alpha^{3}g\,,\qquad
f\mapsto\alpha^{2}f\,. \label{longsctrafo}%
\end{align}
The following combinations of coupling constants are invariant under this
transformation:
\begin{equation}
g/\rho^{3/2}\,,\qquad f/\rho\,. \label{InvCoup}%
\end{equation}
A further symmetry of the Hamiltonian $\mathcal{H}$ is the $z$-dependent field shift
\begin{align}
s(\mathbf{r})  &  \mapsto s(\mathbf{r})+az\,,\nonumber\\
\tau &  \mapsto\tau-\frac{1}{3}ag\,. \label{longtilt}%
\end{align}
The Galilean transformation with transversal $\mathbf{v}$
\begin{align}
s(\mathbf{r})  &  \mapsto s(\mathbf{r-v}z)+\lambda\mathbf{v}\cdot
\mathbf{r\,,}\nonumber\\
\tau &  \mapsto\tau-\frac{\rho^{2}}{2}\mathbf{v}^{2}\,, \label{Galtraf}%
\end{align}
is a symmetry if only and only if the conditions
\begin{equation}
\lambda=\frac{g}{f}\,,\qquad f=\frac{g^{2}}{3\rho^{2}} \label{Galsym}%
\end{equation}
hold. Of course, if $g=0$, the Hamiltonian is invariant under the inversion
$s(\mathbf{r})\mapsto-s(\mathbf{r})$.

\section{Renormalized Perturbation Theory}
\label{sec:RPT} 

In this section, we outline our renormalized perturbation
theory. For background on the methods that we use, we refer to the
textbooks~\cite{ZiJu02,Am84}.

\subsection{Diagrammatic perturbation calculation}

\begin{figure}[ptb]
\centering{\includegraphics[width=4cm]{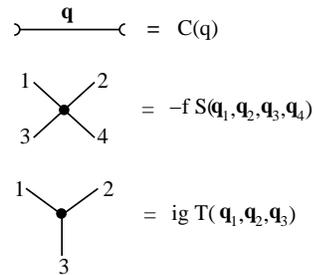}}\caption{Elements of our
perturbation theory: the Gaussian propagator $C (\mathbf{q})$ and the vertices $-fS(\mathbf{q}_{1},\mathbf{q}_{2},\mathbf{q}_{3},\mathbf{q}_{4})$ and $igT(\mathbf{q}_{1},\mathbf{q}_{2},\mathbf{q}_{3})$.}%
\label{fig:StatElem}%
\end{figure}

The Feyman diagrams of our perturbation theory consist of the Gaussian
propagator
\begin{equation}
C(\mathbf{q})=\frac{1}{q_{\bot}^{4}+\rho^{2}q_{\Vert}^{2}+\tau q_{\bot}^{2}%
}=\frac{1}{\bigl(q_{\bot}^{2}+\tau/2\bigr)^{2}+\rho^{2}q_{\Vert}^{2}-\tau
^{2}/4}\,. \label{Prop}%
\end{equation}
and two vertices, see Fig.~\ref{fig:StatElem}. The $\tau^{2}/4$-term in the
denominator of the propagator will be omitted in the following because an
expansion of any diagram in powers of this term will inevitably lead to
unnecessary contributions that are convergent in the ultraviolet (UV). There
is a 3-leg vertex
\begin{equation}
igT(\mathbf{q}_{1},\mathbf{q}_{2},\mathbf{q}_{3})=\frac{ig}{3}\Big[q_{1,\Vert
}\bigl(\mathbf{q}_{2,\bot}\cdot\mathbf{q}_{3,\bot}\bigr)+2\text{
perm.}\Big] \label{3-Vert}%
\end{equation}
and a 4-leg vertex
\begin{align}
-fS(\mathbf{q}_{1},\mathbf{q}_{2},\mathbf{q}_{3},\mathbf{q}_{4})  &
=-\frac{f}{3}\Big[\bigl(\mathbf{q}_{1,\bot}\cdot\mathbf{q}_{2,\bot
}\bigr)\bigl(\mathbf{q}_{3,\bot}\cdot\mathbf{q}_{4,\bot}\bigr)\nonumber\\
&  + 2 \, \, \text{perm.}\Big]\,. \label{4-Vertex}%
\end{align}

We use dimensional regularization about $d=3-\varepsilon$ dimensions in
conjunction with minimal subtraction, i.e., we use minimal renormalization. Our renormalization scheme reads
\begin{align}
s  &  \mapsto\mathring{s}=Z^{1/2}s\,,\qquad\rho^{2}\mapsto\mathring{\rho}%
^{2}=Z^{-1}Z_{\rho}\rho^{2}\,,\nonumber\\
\tau &  \mapsto\mathring{\tau}=Z^{-1}Z_{\tau}\tau+\mathring{\tau}_{c}\,,\qquad
g\mapsto\mathring{g}=Z^{-3/2}Z_{v}g\,,\nonumber\\
f  &  \mapsto\mathring{f}=Z^{-2}\Big(Z_{u}f+Yg^{2}/\rho^{2}\Big)\,,
\label{RenSch}%
\end{align}
where
\begin{equation}
A_{\varepsilon}^{1/2}g=\rho^{3/2}v\mu^{\varepsilon/2}\,,\quad A_{\varepsilon
}f=\rho u\mu^{\varepsilon}\,,\quad A_{\varepsilon}=\frac{\Gamma(1+\varepsilon
/2)}{(4\pi)^{1-\varepsilon/2}}\,.
\end{equation}
We choose the additive renormalization $Y$ so that $Y(u,v)=Y(v)$. The
longitudinal tilt-transformation (\ref{longtilt}) leads to the exact relation%
\begin{equation}
Z_{\tau}=Z_{v}\,. \label{tiltBez}%
\end{equation}
Note that formally $\mathring{\tau}_{c}=0$ in minimal renormalization. In general, i.e., beyond minimal renormalization, $\mathring{\tau}_{c}$ is a positive or negative constant. In a momentum shell RG, for example, it depends on the momentum cut-off and on the coupling constants.

We calculate the 1-loop contributions to the superficially UV-divergent vertex
functions $\Gamma_{2}$, $\Gamma_{3}$ and $\Gamma_{4}$. Some details of this
calculation can be found in Appendix~\ref{app:calcDiagrams1Loop}. For the
counter-terms that render their respective vertex function finite at 1-loop
order, we obtain
\begin{align}
\mathfrak{C}(2)  &  =\frac{v^{2}}{36\varepsilon}\Big(\rho^{2}q_{\Vert}^{2}-\frac{1}%
{2}q_{\bot}^{4}\Big)+\Big(\frac{u}{6\varepsilon}-\frac{v^{2}}{36\varepsilon
}\Big)\tau q_{\bot}^{2}\,,\nonumber\\
\mathfrak{C}(3)  &  =\Big(\frac{v^{2}}{36\varepsilon}-\frac{u}{6\varepsilon
}\Big)\,igT(\mathbf{q}_{1},\mathbf{q}_{2},\mathbf{q}_{3})\,,\nonumber\\
\mathfrak{C}(4)  &  =\Big\{\Big(\frac{3u}{8\varepsilon}-\frac{5v^{2}}{36\varepsilon
}\Big)\,\,f+\frac{v^{2}}{72\varepsilon}\,\frac{g^{2}}{\rho^{2}}%
\Big\}\,S(\mathbf{q}_{1},\mathbf{q}_{2},\mathbf{q}_{3},\mathbf{q}_{4})\,.
\label{Count}%
\end{align}
Form these counter-terms, we extract the 1-loop renomalization factors
\begin{align}
Z  &  =1-\frac{v^{2}}{72\varepsilon}\,,\qquad Z_{\rho}=1+\frac{v^{2}%
}{36\varepsilon}\,,\nonumber\\
Z_{v}  &  =Z_{\tau}=1+\frac{u}{6\varepsilon}-\frac{v^{2}}{36\varepsilon
}\,,\nonumber\\
Z_{u}  &  =1+\frac{3u}{8\varepsilon}-\frac{5v^{2}}{36\varepsilon}\,,\qquad
Y=\frac{v^{2}}{72\varepsilon}\,. \label{RenFakt}%
\end{align}
Note that if the system is invariant under the Galilean
transformation~(\ref{Galtraf}) which implies $v^{2}=3u$, the conditions
(\ref{Galsym}) lead to the following two \emph{exact} relations between the
renormalization factors:
\begin{equation}
Z_{\rho}=Z_{v}=Z_{u}+3Y\,. \label{GalBez}%
\end{equation}

\subsection{Renormalization group and fixed points}

Now, we set up our RG equations and discuss
the RG flow. To simplify the involved equations, we use in the following the
shorthands
\begin{equation}
x:=\frac{v^{2}}{18}\,,\qquad y:=\frac{3u-v^{2}}{18} \label{x-y}%
\end{equation}
for the renormalized versions of the invariant combination of coupling
constants given in Eq.~(\ref{InvCoup}). In terms of $x$ and $y$, the RG
$\gamma$-functions are defined as
\begin{align}
\gamma_{\cdots}  &  =\left.  \mu\partial_{\mu}\ln Z_{\cdots}\right\vert
_{0}=-(x\partial_{x}+y\partial_{y})Z_{\cdots}^{(1)}\,,\nonumber\\
\alpha &  =(x\partial_{x}+y\partial_{y})Y^{(1)}\,,
\end{align}
where $Z_{..}^{(1)}$ and $Y^{(1)}$ denote the $\varepsilon$-residues of the
respective renormalization factors. To 1-loop order, we obtain
\begin{align}
\gamma &  =\frac{x}{4}\,,\qquad\gamma_{\rho}=-\frac{x}{2}\,,\qquad\gamma
_{\tau}=\gamma_{v}=-\frac{x}{2}-y\,,\nonumber\\
\gamma_{u}  &  =\frac{x-9y}{4}\,,\qquad\alpha=\frac{x}{4}\,. \label{WilsFu}%
\end{align}
Inspecting the relations (\ref{GalBez}), in particular for $y=0$,  we figure that it is useful
to define
\begin{equation}
\sigma=\bigl(3\alpha-\gamma_{u}+2\gamma_{v}-\gamma_{\rho}\bigr)/y\,.
\label{sigma-def}%
\end{equation}
This quantity has a perturbation expansion in $x$ and $y$ which is to lowest order given by
\begin{equation}
\sigma=\frac{1}{4}\,. \label{sigma}%
\end{equation}
From the RG $\gamma$-functions, we get the Wilson and Gell-Mann--Low
functions
\begin{align}
\zeta &  =\left.  \mu\partial_{\mu}\ln\rho\right\vert _{0}=\frac{\gamma
-\gamma_{\rho}}{2}\,,\nonumber\\
\kappa &  =\left.  \mu\partial_{\mu}\ln\tau\right\vert _{0}=\gamma
-\gamma_{\tau}\,,\nonumber\\
\beta_{x}  &  =\left.  \mu\partial_{\mu}x\right\vert _{0}=\Big(-\varepsilon
+\frac{3}{2}\gamma+\frac{3}{2}\gamma_{\rho}-2\gamma_{v}\Big)x\,,\nonumber\\
\beta_{y}  &  =\left.  \mu\partial_{\mu}y\right\vert _{0}=\Big(-\varepsilon
+\frac{3}{2}\gamma+\frac{1}{2}\gamma_{\rho}-\gamma_{u}+\sigma x\Big)y\,.
\label{RGFu}%
\end{align}
Note that for $y=0$, which corresponds to the GP model where the Ward
identities (\ref{GalBez}) do hold, we have
\begin{equation}
\gamma_{\rho}=\gamma_{v}
\label{GammaGalBez}%
\end{equation}
to arbitrary order in perturbation theory. Our 1-loop results are consistent
with these relations and hence satisfy an important check.
Having these RG functions, we can write down the RG differential operator. Applying
 $\left. \mu\partial_{\mu}\cdots\right\vert _{0}$ to cumulants (Green functions) amounts to to RG equation (RGE)
\begin{equation}
\mathcal{D}_{\mu}s(\mathbf{r})=-\frac{\gamma}{2}s(\mathbf{r})\,, \label{RGGr}%
\end{equation}
with the RG differential operator
\begin{equation}
\mathcal{D}_{\mu}:=\mu\partial_{\mu}+\zeta\rho\partial_{\rho}+\kappa
\tau\partial_{\tau}+\beta_{x}\partial_{x}+\beta_{y}\partial_{y}\,.
\label{RGOp}%
\end{equation}

Now, we are in position to discuss the fixed points of the RG. After setting $\mu\rightarrow\mu\ell$, the flow of $x$ and $y$ is given 
by the ordinary differential equations
\begin{equation}
\ell\frac{d\bar{x}(\ell)}{d\ell}=\beta_{x}(\bar{x}(\ell),\bar{y}%
(\ell))\,,\quad\ell\frac{d\bar{y}(\ell)}{d\ell}=\beta_{y}(\bar{x}(\ell
),\bar{y}(\ell))\, \label{flow}%
\end{equation}
with $\bar{x}(1)=x$ and $\bar{y}(1)=y$. These fixed points, which are approached
asymptotically for $\ell\rightarrow0$, follow from the zeros of the
Gell-Mann--Low functions,
\begin{equation}
\left.  \beta_{x}\right\vert _{\ast}=\left.  \beta_{y}\right\vert _{\ast}=0\,.
\label{RGFP}%
\end{equation}
with these functions given to $1$-loop order by
\begin{align}
\beta_{x}  &  =\Big(-\varepsilon+\frac{5}{8}x+2y\Big)x\,,\nonumber\\
\beta_{y}  &  =\Big(-\varepsilon+\frac{1}{8}x+\frac{9}{4}y\Big)y\,.
\label{beta1L}%
\end{align}

\subsubsection{$\varepsilon>0$}
In addition to the unstable trivial fixed point $x_{\ast}^{(0)}=y_{\ast}%
^{(0)}=0$, there are three more fixed points for $\varepsilon>0$.
(1) The Galilean-invariant fixed point
\begin{equation}
x_{\ast}^{(1)}=\frac{8}{5}\varepsilon+O(\varepsilon^{2})\,,\qquad y_{\ast
}^{(1)}=0\,, \label{FPGal}%
\end{equation}
that corresponds to the GP model, $y=0$. We will refer to this fixed point as the GP fixed point. (2) The inversion-invariant fixed
point
\begin{equation}
x_{\ast}^{(2)}=0\,,\qquad y_{\ast}^{(2)}=\frac{4}{9}\varepsilon+O(\varepsilon
^{2})\,, \label{FPSch}%
\end{equation}
that corresponds to the FIS model. We will refer to this fixed point as the FIS fixed point. (3) The new fixed point
\begin{equation}
x_{\ast}^{(3)}=\frac{8}{37}\varepsilon+O(\varepsilon^{2})\,,\qquad y_{\ast
}^{(3)}=\frac{16}{37}\varepsilon+O(\varepsilon^{2})\,, \label{FPneu}%
\end{equation}
We will see below that this is the only stable fixed point.

Next, we discuss the stability of the fixed points which is determined by the
stability matrix whose components are given by the partial derivatives with
respect to $x$ and $y$, respectively,
\begin{equation}
\underline{\underline{\beta}}=%
\begin{pmatrix}
-\varepsilon+\frac{5}{4}x+2y\,, & 2x\\
\frac{1}{8}y\,, & -\varepsilon+\frac{1}{8}x+\frac{9}{2}y
\end{pmatrix}
\,. \label{StabMat}%
\end{equation}
Its eigenvalues (the Wegner correction exponents) and the corresponding right
eigenvectors which indicate the flow near the respective fixed points are as
follows: (1) For the GP fixed point, we have $\omega_{1}%
^{(1)}=\varepsilon$ with $\underline{e}_{1}^{(1)}=(1,0)$ and $\omega_{2}%
^{(1)}=-4\varepsilon/5$ with $\underline{e}_{2}^{(1)}=(-1,9/16)$. Thus, this
fixed point is stable only in one direction. (2) For the FIS
fixed point, we have $\omega_{1}^{(2)}=\varepsilon$ with $\underline{e}%
_{1}^{(2)}=(0,1)$ and $\omega_{2}^{(2)}=-\varepsilon/9$ with $\underline
{e}_{2}^{(2)}=(1,-1/20)$. Hence, this fixed point is also stable in only one
direction. (3) For the new fixed point, we have $\omega_{1}^{(3)}=\varepsilon$
with $\underline{e}_{1}^{(3)}=(1/2,1)$ and $\omega_{2}^{(3)}=4\varepsilon/37$
with $\underline{e}_{2}^{(3)}=(1,-1/16)$. Therefore, the new fixed point is
stable in any direction.

\begin{figure}[ptb]
\centering{\includegraphics[width=5cm]{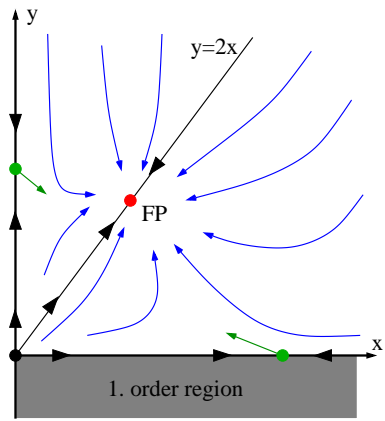}}\caption{(Color online) Renormalization
group flow for $\varepsilon >0$. The (black) dot in the origin indicates the unstable Gaussian fixed point. The (green) dot on the abscissa marks the unstable GP fixed point, and the (green) dot on the ordinate marks the unstable FIS fixed point. The (red) dot on the line $y=2x$ represents the stable fixed point of our model. Arrows indicate the direction of the RG flow.}%
\label{fig:flussbild}%
\end{figure}

The picture of the RG flow that emerges from {the three invariant lines }%
$x=0$, $y=0$, and $y=2x$, as well as from the fixed points and their
respective stability is sketched in Fig.~(\ref{fig:flussbild}) and can be
described as follows: The domain of attraction of the stable fixed point is
given by the positive quadrant $x\geq0$, $y\geq0$, that is bounded by the
lines $y=0$ (GP model) and $x=0$ (FIS model). The unstable fixed points that
correspond to the GP and FIS models, respectively, are located on these lines.
For negative $y$, the RG flow runs off to infinity. This suggests that we are
dealing with a fluctuation-induced first order transition in the case
$v^{2}>u$. 

\subsubsection{$\varepsilon\leq0$}
Of course, the trivial fixed point $x_{\ast}^{(0)}=y_{\ast}^{(0)}=0$ is stable
for $\varepsilon\leq0$. Thus, for dimensions $d>3$, we find mean-field
behavior. Right at the critical dimension, the mean-field power laws have logarithmic corrections. This case is of particular importance here because for the elastic phase transitions the upper critical dimension and the physical dimension are the same. In the next section, we will discuss these logarithmic corrections in detail. As a prelude, we study here the asymptotic flow of $x$ and $y$ to zero for $\varepsilon = 0$.

For $\varepsilon = 0$, the flow equations (\ref{flow}) in conjunction with Gell-Mann--Low functions
(\ref{beta1L}) lead to the differential equations%
\begin{align}
\ell\frac{d\bar{x}(\ell)}{d\ell} &  =\Big(\frac{5}{8}\bar{x}(\ell)+2\bar
{y}(\ell)\Big)\bar{x}(\ell)\,,\nonumber\\
\ell\frac{d\bar{y}(\ell)}{d\ell} &  =\Big(\frac{1}{8}\bar{x}(\ell)+\frac{9}%
{4}\bar{y}(\ell)\Big)\bar{y}(\ell)\,\label{floweq}%
\end{align}
Solving these equations for $\ell\ll1$, we obtain%
\begin{align}
\bar{x}(\ell) &  =\frac{a}{\left\vert \ln\ell\right\vert }+O(\left\vert
\ln\ell\right\vert ^{-2})\,,\nonumber\\
\bar{y}(\ell) &  =\frac{b}{\left\vert \ln\ell\right\vert }+O(\left\vert
\ln\ell\right\vert ^{-2})\,.\label{asympt}%
\end{align}
The amplitudes are given in general by
\begin{equation}
a=\frac{8}{37}\,,\qquad b=\frac{16}{37}\,.\label{ampl}%
\end{equation}
The special models lead to%
\begin{align}
\text{GP} &  \text{:}\qquad b=0\,,\qquad a=\frac{8}{5}\,,\nonumber\\
\text{FIS} &  \text{:}\qquad a=0\,,\qquad b=\frac{4}{9}\,.\label{specampl}%
\end{align}

\section{Scaling and critical exponents}
\label{sec:scalingAndExponents}
In this section, we first study the scaling behavior of the cumulants of our theory. Then, we extract from this general result the critical behavior of several experimentally relevant quantities for driven surface diffusion and elastic transitions, respectively.

\subsection{General scaling behavior}
Now, we determine scaling properties of the cumulants $G_{n,m}(\{\mathbf{r}\},\tau)$. Besides the
order-parameter-field $s$ and its spatial derivatives, we consider the
\textquotedblleft energy\textquotedblright-field $e=(\nabla_{\bot}s)^{2}/2$.
This field is a composite field in the language of renormalization theory with
its own scaling behavior. This scaling behavior can be studied via so-called insertions that are generated by 
variations $\delta\cdots/\delta\tau(\mathbf{r})$ of the cumulants with respect to the local \textquotedblleft temperature\textquotedblright $\tau (\mathbf{r})$. 
In other words, the objects of our study here are the expectation values 
\begin{equation}
G_{n,m}(\{\mathbf{r}\},\tau)=\langle\lbrack s]^{n}[e]^{m}\rangle
^{(cum)}=\Big[-\frac{\delta}{\delta\tau}\Big]^{m}G_{n,0} \, ,
\label{kum}%
\end{equation}
which we have written here in a somewhat compressed yet evident form.

To extract the scaling properties of the Green functions, we first note that simple dimensional analysis gives
\begin{align}
&  G_{n,m}(\{z,\mathbf{r}_{\bot}\},\tau;x,y;\rho,\mu)\nonumber\\
&  =\rho^{-n/2-m}G_{n,m}(\{\rho^{-1}z,\mathbf{r}_{\bot}\},\tau;x,y;1,\mu
)\nonumber\\
&  =\mu^{n(d-3)/2+m(d-1)}G_{n,m}(\{\mu^{2}z,\mu\mathbf{r}_{\bot}\},\mu
^{-2}\tau;x,y;\rho,1)
 \label{kumscal}%
\end{align}
for the leading, mean-field behavior. Next, we solve the RGE for the Green functions to get the anomalous contributions to scaling. For $n\geq1$, this RGE reads
\begin{align}
\bigl(\mu\partial_{\mu}+\zeta\rho\partial_{\rho}+\kappa\tau\partial_{\tau}  &
+\beta_{x}\partial_{x}+\beta_{y}\partial_{y}\nonumber\\
&  +\frac{n}{2}\gamma+m\kappa\bigr)G_{n,m}=0\,.
 \label{RGGl}%
\end{align}
As usual, this partial differential
equation can be solved by the method of characteristics. The characteristics read $\bar{\mu}(\ell
)=\mu\ell$, $\bar{\rho}(\ell)=X_{\rho}(\ell)\rho$ and $\bar{\tau}(\ell)=X_{\tau
}(\ell)\tau$, where the $X$-factors are the solutions of the ordinary
differential equations%
\begin{align}
&  \ell\frac{d\ln X(\ell)}{d\ell}=\gamma(\bar{x}(\ell),\bar{y}(\ell))\,,\qquad
X(1)=1\,,\nonumber\\
&  \ell\frac{d\ln X_{\rho}(\ell)}{d\ell}=\zeta(\bar{x}(\ell),\bar{y}%
(\ell))\,,\qquad X_{\rho}(1)=1\,,\nonumber\\
&  \ell\frac{d\ln X_{\tau}(\ell)}{d\ell}=\kappa(\bar{x}(\ell),\bar{y}%
(\ell))\,,\qquad X_{\tau}(1)=1\,. \label{ampldiff}%
\end{align}
Collecting the solutions of these equations, the flowing couplings $\bar{x}(\ell),\bar{y}(\ell)$ that solve
Eq.~(\ref{flow}) and the mean-field scaling as given in  Eq.~(\ref{kumscal}), we obtain the scalings forms
\begin{align}
&  G_{n,m}(\{z,\mathbf{r}_{\bot}\},\tau;x,y;\rho,\mu)\nonumber\\
&  =\Big(\ell^{d-3}X_{\rho}(\ell)^{-1}X(\ell)\Big)^{n/2}\Big(\ell^{d-1}%
X_{\rho}(\ell)^{-1}X_{\tau}(\ell)\Big)^{m}\nonumber\\
&  \times G_{n,m}\bigl(\{\ell^{2}X_{\rho}(\ell)^{-1}z,\ell\mathbf{r}_{\bot
}\},\ell^{-2}X_{\tau}(\ell)\tau;\bar{x}(\ell),\bar{y}(\ell);\rho,\mu\bigr)\,.
\label{Genscal}%
\end{align}
This result is correct as long as $n\geq1$. 

In the case $n=0$, $m=1$ or $2$ one has
to account for additive renormalizations which become essential for the specific
heat for $\varepsilon=0$. The additive renormalizations and its consequences
are the theme of Appendix~\ref{app:addRen}. We apply the scalings
(\ref{kumscal}) to the result (\ref{G-fin}) and obtain%
\begin{align}
&  G_{0,m}(\{z,\mathbf{r}_{\bot}\mathbf{\},\tau};x,y;\rho,\mu)=\Big((\ell
\mu)^{2}(X_{\rho}(\ell)\rho)^{-1}X_{\tau}(\ell)\Big)^{m}\nonumber\\
&  \times G_{0,m}\bigl(\{(\ell\mu)^{2}(X_{\rho}(\ell)\rho)^{-1}z,\ell
\mu\mathbf{r}_{\bot}\},(\ell\mu)^{-2}X_{\tau}(\ell)\tau;\nonumber\\
&  \qquad\qquad\qquad\qquad\qquad\qquad\qquad\qquad\qquad\bar{x}(\ell),\bar
{y}(\ell);1,1\bigr)\nonumber\\
&  \qquad\qquad+\frac{(-\tau)^{2-m}}{\rho}I(\ell)[\delta(\mathbf{r}%
_{1}-\mathbf{r}_{2})]^{m-1}\,, \label{enkorr}%
\end{align}
with $I(\ell)$ given by Eq.~(\ref{I-l}).

\subsection{Driven surface diffusion}

In experiments on surface growth, one often measures the roughness of a surface and the associated roughness exponents. These exponents are defined through the height-height correlation function%
\begin{equation}
C(z,\mathbf{r}_{\bot},\tau)=\langle\bigl(h(z,\mathbf{r}_{\bot})-h(0,\mathbf{0}%
)\bigr)^{2}\rangle\,. \label{hh-corr}%
\end{equation}
For surface diffusion, the physical dimension is $d=2$.
Hence, we are interested in the scaling behavior of the height-height
correlation and its roughening properties for $\varepsilon>0$. The asymptotic
solutions of the RG-equations (\ref{ampldiff}) are%
\begin{align}
X(\ell)  &  =\ell^{\eta}\,,\qquad\eta=\gamma_{\ast}\,,\nonumber\\
X_{\rho}(\ell)  &  =\ell^{\zeta_{\ast}}\,,\qquad X_{\tau}(\ell)=\ell
^{\kappa_{\ast}}\,, \label{amplscal}%
\end{align}
where $\gamma_{\ast}$, $\zeta_{\ast}$, and $\kappa_{\ast}$ are the values of
the RG-functions at the fixed point $x_{\ast}$, $y_{\ast}$. For the stable
fixed point as well as for the GP fixed point, we have $x_{\ast}>0$. Hence we get from the definitions (\ref{RGFu})
the exact relation%
\begin{equation}
2\kappa_{\ast}-3\zeta_{\ast}=\varepsilon-\eta\,. \label{exrel}%
\end{equation}
The scaling behavior of the height-height correlation function~(\ref{hh-corr}) can be deduced from the general result
(\ref{Genscal}):
\begin{equation}
C(z,\mathbf{r}_{\bot},\tau)=\ell^{d-4+\Delta+\eta}C\bigl(\ell^{1+\Delta}%
z,\ell\mathbf{r}_{\bot},\ell^{-1/\nu}\tau\bigr)\,, \label{hh-scal}%
\end{equation}
where we have defined the anisotropy and the correlation length exponents%
\begin{align}
\Delta &  =1-\zeta_{\ast}\,,\nonumber\\
\nu &  =\frac{1}{2-\kappa_{\ast}}\,, \label{Delt-nu}%
\end{align}
which are related (with exception of the FIS-model) by%
\begin{equation}
d-2+\eta=\frac{2}{\nu}-3\Delta\,.
\end{equation}
Due to the presence of strong anisotropy, we have to dealing with two different
roughness exponents $\alpha_{\bot}$ and $\alpha_{\Vert}$ defined by
\begin{subequations}
\begin{align}
C(0,\mathbf{r}_{\bot},\tau)  &  \equiv\left\vert \mathbf{r}_{\bot}\right\vert
^{2\alpha_{\bot}}c_{\bot}\bigl(\tau\left\vert \mathbf{r}_{\perp}\right\vert
^{1/\nu}\bigr)\,,\nonumber\\
C(z,\mathbf{0},\tau)  &  \equiv\left\vert z\right\vert ^{2\alpha_{\Vert}%
}c_{_{\Vert}}\bigl(\tau z^{1/(1+\Delta)\nu}\bigr)\,. \label{roughdef}%
\end{align}
The two exponents
\end{subequations}
\begin{align}
\alpha_{\bot}  &  =\frac{1}{2}(4-d-\eta-\Delta)=1+\Delta-1/\nu\text{
},\nonumber\\
\alpha_{\Vert}  &  =\frac{\alpha_{\bot}}{1+\Delta}=1-\frac{1}{(1+\Delta)\nu
}\,, \label{roughexp}%
\end{align}
are read off immediately. In the dimensional-expansion about three
surface-dimensions, we get%
\begin{align}
\Delta &  =1-\frac{3\varepsilon}{37}\,,\qquad\frac{1}{\nu}=2-\frac
{22\varepsilon}{37}\,,\nonumber\\
\alpha_{\bot}  &  =\frac{19\varepsilon}{37}\,,\qquad\alpha_{\Vert}%
=\frac{19\varepsilon}{74}\,,\qquad\eta=\frac{2\varepsilon}{37}\,,
\label{epsexp}%
\end{align}
to first order in $\varepsilon$.

\subsection{Elastic phase transition}
As mentioned above, the physical and the upper critical dimensions coincide for elastic transitions.
Therefore we have $\varepsilon=0$, and we are interested in logarithmic corrections. The typical measurable
quantities are the susceptibility 
\begin{align}
\chi &  =\int dzd^{2}r_{\bot}\,\langle\nabla_{\bot}u(z,\mathbf{r}_{\bot}%
)\cdot\nabla_{\bot}u(0,\mathbf{0})\rangle\nonumber\\
&  =\int dzd^{2}r_{\bot}\,\bigl(-\nabla_{\bot}^{2}\bigr)G_{2,0}(z,\mathbf{r}%
_{\bot},\tau)
\label{DefSus}
\end{align}
 of the critical modes $\nabla_{\bot}u$, and the critical part of the specific heat
\begin{align}
c &  =\int dzd^{2}r_{\bot}\langle e(z,\mathbf{r}_{\bot})e(0,\mathbf{0}%
)\rangle\nonumber\\
&  =\int dzd^{2}r_{\bot}\,G_{0,2}(z,\mathbf{r}_{\bot},\tau
)\,.\label{DefSpez}%
\end{align}
that, as the second line Eq.~(\ref{DefSpez}) indicates, can be calculated from the energy correlation function.
Here, we need the asymptotic solutions of the RG-equations (\ref{ampldiff}) for the
amplitudes using the assymtotic flow (\ref{asympt}). We get%
\begin{align}
X(\ell) &  \simeq\left\vert \ln\ell\right\vert ^{-a/4}\,,\qquad X_{\rho}%
(\ell)\simeq\left\vert \ln\ell\right\vert ^{-3a/8}\,,\nonumber\\
X_{\tau}(\ell) &  \simeq\left\vert \ln\ell\right\vert ^{-3a/4-b}%
\,,\label{AmplLog}%
\end{align}
with the constants $a$ and $b$  given in Eqs.~(\ref{ampl}) and (\ref{specampl}).
From the general scaling properties of the Greens-functions (\ref{Genscal})
and bysetting%
\begin{equation}
\ell^{2}=X_{\tau}(\ell)\left\vert \tau\right\vert \,,\label{l-tau}%
\end{equation}
(we have set the scale so that $\mu,\rho\rightarrow1$ to simplify the
conclusions), we obtain 
\begin{align}
\chi(\tau) &  =\ell^{-2}X(\ell)\chi(\ell^{-2}X_{\tau}(\ell)\tau)\nonumber\\
&  =\frac{X(\ell)}{X_{\tau}(\ell)\left\vert \tau\right\vert }\chi
(1)\simeq\frac{\left\vert \ln\left\vert \tau\right\vert \right\vert
^{\gamma_{0}}}{\left\vert \tau\right\vert }\label{SusLog}%
\end{align}
asymptotically. The correction exponent $\gamma_{0}=a/2+b$ reads
\begin{equation}
\gamma_{0}=\frac{20}{37}\,,
\end{equation}
whereas the special models lead to%
\begin{align}
\text{GP} &  \text{:}\qquad\gamma_{0}=\frac{4}{5}\,,\nonumber\\
\text{FIS} &  \text{:}\qquad\gamma_{0}=\frac{4}{9}\,.
\end{align}
Next, using the scaling result for the energy correlations (\ref{enkorr}) we get
\begin{align}
c(\tau) &  =\frac{X_{\tau}(\ell)^{2}}{X_{\rho}(\ell)}c(\ell^{-2}X_{\tau}%
(\ell)\tau)+\frac{1}{\rho}I(\ell)\nonumber\\
&  =\frac{X_{\tau}(\ell)^{2}}{X_{\rho}(\ell)}c(1)+\frac{1}{\rho}%
I(\ell)\nonumber\\
&  \simeq\text{const.}\left\vert \ln\left\vert \tau\right\vert \right\vert
^{\alpha_{0}-1}+\frac{\left\vert \ln\left\vert \tau\right\vert \right\vert
^{\alpha_{0}}}{\alpha_{0}}\,.\label{SpezLog}%
\end{align}
with the correction exponent $\alpha_{0}=1-9a/8-2b$. We recognize that the term
originating from the additive renormalization overwhelms the first one
asymptotically. The correction exponent reads
\begin{equation}
\alpha_{0}=-\frac{4}{37}\,.
\end{equation}
The special models lead to%
\begin{align}
\text{GP} &  \text{:}\qquad\alpha_{0}=-\frac{4}{5}\,,\nonumber\\
\text{FIS} &  \text{:}\qquad\alpha_{0}=\frac{1}{9}\,.
\end{align}

Note that our results for the correction exponents for the susceptibility and the specific heat of the FIS model are in full agreement with those obtained by FIS originally. Hence, our calculation passes an another consistency check.


\section{Concluding Remarks and Outlook}
\label{sec:concludingRemarks}

In summary, we have studied driven surface diffusion at the phase transition from isotropic, Edwards-Wilkinson type behavior to a state characterized by longitudinal ripples and continuous elastic phase transitions. We have shown that the two transitions belong under appropriate conditions to the same universality class. We derived the field theoretic Hamiltonian for this universality class which has an upper critical dimension of $d_c =3$. The dimension $d$ plays different roles in the two systems. In driven surface diffusion, $d$ is the dimension of the substrate whereas for the elastic phase transition, $d$ is the full dimension of space. Hence, physically, $d=2$ in the former problem and $d=3$ in the latter. Thus, the critical behavior in the surface diffusion problem is described by power laws with anomalous scaling exponents whereas that of the elastic phase transitions under consideration is described by power laws with mean-field exponents decorated by logarithmic corrections. We calculated these anomalous exponents and logarithmic corrections, respectively, for several experimentally relevant quantities. 

As far as elastic phase transitions were concerned, we restricted ourselves to crystals with tetragonal or higher symmetry, i.e., crystals whose harmonic Lagrange energy density can be
obtained from the tetragonal case by imposing relations between certain
elastic constants because crystals with lower symmetry such as orthorhombic or monoclinic symmetry have simple mean field behavior. In addition to that, we have left aside crystals with
cubic symmetry in the transversal subspace. These do have anomalous scaling behavior, however, the diagrammatic perturbation calculation gets significantly more involved when cubic terms are present and is beyond the scope or the present paper. We will present our results for the cubic case in a future publication~\cite{JanSteUnpub}.

Our model Hamiltonian differs from the SPJ model in that we do not assume tilt invariance and that we do assume detailed balance, at least on a semi-macroscopic scale. Our result for the anisotropy exponent $\Delta$ is smaller and those for the roughness exponents are larger than the corresponding 1-loop results produced by the SPJ model. Hence, in comparison, our model leads to a less anisotropic but slightly rougher surface. Its restriction to detailed balance makes our model a special case of a more general model. Our results, moreover, show that our model corresponds to a RG fixed point of this more general model. Whether or not this fixed point is the stable fixed point of the more general model, we cannot judge at this point, and we will address this question in future work. This question is also important from an experimental standpoint. If our fixed point turns out being stable, this would mean that the asymptotic behavior in a typical experiment on driven surface diffusion with oblique incidence would show detailed balance as characterized by our fixed point.

Our model Hamiltonian differs from the FIS model in that it includes a relevant third order coupling that does not destroy the continuous character of the transition. The presence of this term modifies the exponents of the logarithmic corrections to the susceptibility and the specific heat. For the former, the difference is quantitative, for the latter it is even qualitative. The FIS model predicts that the specific heat diverges as the transition is approached, whereas we find that it vanishes. As the FIS model, our model entirely neglects the noncritical transversal components of the elastic displacement field. If the order parameter field corresponded directly to a component of the strain tensor, one could simply integrate out the noncritical strain components. This would produce additional negative contributions to the positive (on grounds of stability) fourth order coupling constant of our model. If the additional negative contributions exceed the positive contribution in magnitude, the transition becomes first order. However, here we have to integrate out the transversal displacement components and not entire components of the strain tensor. The transversal components of the displacement are contained together with its critical components in different components of the strain tensor, and it is not clear how to integrate them out. We will return to the question of the influence of the noncritical displacements in future work.

From a conceptual standpoint, it is worthwhile to reflect on the different roles played by reference and target space symmetries in elastic phase transitions and related problems~\cite{tomPrivate}. Invariance under rigid rotations in target space, unless broken by external fields, is a symmetry that applies likewise to any crystal, or for that matter to any elastic system that can be described by Lagrange elasticity theory, including smectic liquid crystals, liquid crystal elastomers etc. Reference space symmetry, on the other hand, is determined by the crystallographic point group of a given elastic system. It has been an established approach for setting up field theoretic models for elastic systems for about three decades to employ gradient expansion and then to retain only those terms that are deemed relevant based on dimensional analysis. Examples where this approach has been used include, to name a few, the original work on nematic liquid crystals by GP, isotropic~\cite{aronovitzLub1988} and anisotropic~\cite{radziToner1995} tethered membranes, sliding columnar phases~\cite{oHernLubensky1998}, nematic elastomers~\cite{xingRad2003,stenullLub2003} and their membranes~\cite{xingMkLubRad2003} and so on, and we also used this approach here. For systems with comparatively high reference space symmetry, vestiges of the target space rotational invariance remain in the final field theoretic Hamiltonian. The nonlinear part of the strain $\partial_{\Vert}u+\frac{1}{2}(\nabla_{\bot}u)^{2}$ in the GP model, for example, ensures that the target space symmetry is realized at least linearly. The situation is similar in the other examples just mentioned. For the elastic transition studied in the present paper, with exception of the limiting isotropic case that coincides with the GP model, the reference space symmetry is comparatively low, and there are no remains of the target space rotational invariance in the final Hamiltonian. It transpires that the extend to which the latter symmetry survives when setting up field theories depends on the former and that, in a sense, reference space symmetry is superior to target space symmetry. It is desirable to get a deeper understanding of the interplay of the two symmetries in field theories, and we see here an interesting opportunity for future work.

\begin{acknowledgments}
This work was supported in part (OS) by the National Science Foundation (DMR1104707 and MRSEC DMR-1120901). OS thanks T.C. Lubensky for spurring his interest in elastic phase transitions and useful discussions.
\end{acknowledgments}

\appendix

\section{Calculation of Feynman Diagrams}
\label{app:calcDiagrams1Loop}

\begin{figure}[ptb]
\centering{\includegraphics[width=6cm]{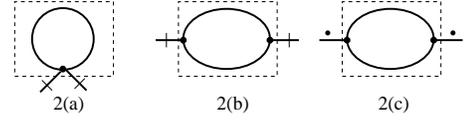}}\caption{2-leg diagrams. A
dash symbolizes a transversal and a point a longitudinal derivative at an
external leg.}%
\label{fig:2-leg}%
\end{figure}
\begin{figure}[ptbptb]
\centering{\includegraphics[width=5cm]{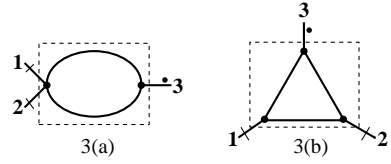}}\caption{3-leg diagrams.}%
\label{fig:3-leg}%
\end{figure}
\begin{figure}[ptbptbptb]
\centering{\includegraphics[width=7cm]{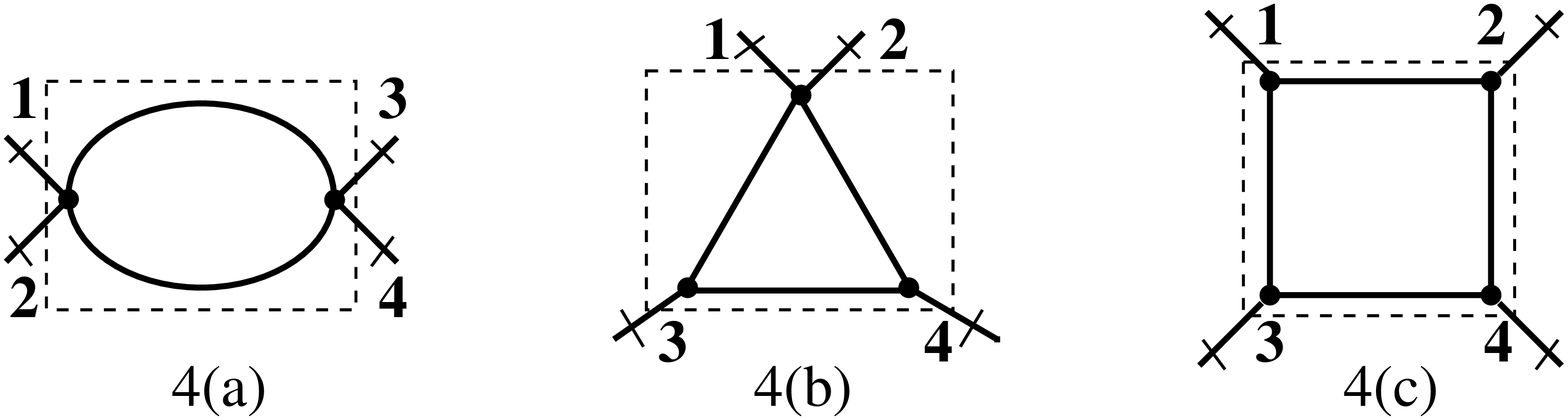}}\caption{4-leg diagrams.}%
\label{fig:4-leg}%
\end{figure}
Here, we present a few cornerstones of our calculation of Feynman diagrams. The 1-loop diagrams of our theory are depicted in Figs.~\ref{fig:2-leg} to \ref{fig:4-leg}. A typical integral encountered in calculating these diagrams reads
\begin{align}
F_{\alpha\beta}=\int_{\mathbf{q}_{1}+\mathbf{q}_{2}=\mathbf{q}}  &
\frac{q_{1,\Vert}q_{1,\bot\alpha}}{\Big(\bigl(q_{1,\bot}^{2}+\tau
/2\bigr)^{2}+\rho^{2}q_{1,\Vert}^{2}\Big)}\nonumber\\
&  \times\,\frac{q_{2,\Vert}q_{2,\bot\beta}}{\Big(\bigl(q_{2,\bot}^{2}%
+\tau/2\bigr)^{2}+\rho^{2}q_{2,\Vert}^{2}\Big)}%
\end{align}
where $\int_{\mathbf{q}}\ldots=(2\pi)^{-d}\int\ldots d^{d}q$. In all of our 1-loop integrals, we can set the external longitudinal momentum to
zero. Via setting $q_{1,\Vert}=-q_{2,\Vert}=\omega/\rho$, the integration over
the internal longitudinal momentum can then be performed using the identity
\begin{equation}
\frac{1}{2\pi}\int_{-\infty}^{\infty}\frac{d\omega}{(\omega^{2}+c^{2})^{n}%
}=\frac{(2(n-1))!}{[(n-1)!]^{2}(2c)^{2n-1}}\,. \label{Hilfe}%
\end{equation}
We obtain
\begin{equation}
F_{\alpha\beta}=-\frac{1}{2\rho^{3}}\int_{\mathbf{q}_{1,\bot}+\mathbf{q}%
_{2,\bot}=\mathbf{q}_{\bot}}\frac{q_{1,\bot\alpha}q_{2,\bot\beta}%
}{\bigl(q_{1,\bot}^{2}+q_{2,\bot}^{2}+\tau\bigr)}\,. \label{IntF1}%
\end{equation}
The remaining integrations are simplified by switching integration variables
to $\mathbf{q}_{1,\bot}=\mathbf{q}_{\bot}/2+\mathbf{p}_{\bot}$ and
$\mathbf{q}_{2,\bot}=\mathbf{q}_{\bot}/2-\mathbf{p}_{\bot}$ and defining
$m^{2}:=\tau/2+q_{\bot}^{2}/4$:
\begin{align}
F_{\alpha\beta}  &  =-\frac{1}{16\rho^{3}}\int_{\mathbf{p}_{\bot}}%
\frac{q_{\bot\alpha}q_{\bot\beta}-4p_{\bot\alpha}p_{\bot\beta}}{p_{\bot}%
^{2}+m^{2}}\nonumber\\
=  &  -\frac{1}{16\rho^{3}}\int_{\mathbf{p}_{\bot}}\frac{q_{\bot\alpha}%
q_{\bot\beta}-4p_{\bot}^{2}\delta_{\alpha\beta}/(d-1)}{p_{\bot}^{2}+m^{2}}
\label{IntF2}%
\end{align}
Using dimensional regularization of the integral with $d=3-\varepsilon$, we
finally obtain
\begin{align}
F_{\alpha\beta}  &  =-\frac{A_{\varepsilon}\mu^{-\varepsilon}}{16\rho
^{3}\varepsilon}\Big\{2q_{\bot\alpha}q_{\bot\beta}+\bigl(q_{\bot}^{2}%
+2\tau\bigr)\delta_{\alpha\beta}\Big\}\nonumber\\
&  \qquad\qquad\qquad\qquad\qquad\qquad+O(\varepsilon^{0})\,, \label{IntF3}%
\end{align}
where $A_{\varepsilon}=\Gamma(1+\varepsilon/2)/(4\pi)^{(d-1)/2}$. All other
Feyman integrals at 1-loop order can be calculated by similar means. We get:%
\begin{align}
G_{\alpha\beta}=  &  \int_{\mathbf{q}_{1}+\mathbf{q}_{2}=\mathbf{q}}%
\frac{q_{1,\bot\alpha}q_{1,\bot\beta}}{\Big(\bigl(q_{1,\bot}^{2}%
+\tau/2\bigr)^{2}+\rho^{2}q_{1,\Vert}^{2}\Big)}\nonumber\\
&  \qquad\times\,\frac{q_{2,\Vert}q_{2,\Vert}}{\Big(\bigl(q_{2,\bot}^{2}%
+\tau/2\bigr)^{2}+\rho^{2}q_{2,\Vert}^{2}\Big)}\nonumber\\
&  =\frac{A_{\varepsilon}\mu^{-\varepsilon}}{16\rho^{3}\varepsilon
}\Big\{2q_{\bot\alpha}q_{\bot\beta}-\bigl(q_{\bot}^{2}+2\tau\bigr)\delta
_{\alpha\beta}\Big\}\nonumber\\
&  \qquad\qquad\qquad\qquad\qquad\qquad+O(\varepsilon^{0})\,. \label{IntG}%
\end{align}%
\begin{align}
H_{\alpha\beta}  &  =\int_{\mathbf{p}}\frac{p_{\bot\alpha}p_{\bot\beta}%
}{\Big(\bigl(p_{\bot}^{2}+\tau/2\bigr)^{2}+\rho^{2}p_{\Vert}^{2}%
\Big)}\nonumber\\
&  =-\frac{A_{\varepsilon}\mu^{-\varepsilon}}{4\rho\varepsilon}\tau
\delta_{\alpha\beta}+O(\varepsilon^{0})\,. \label{IntH}%
\end{align}%
\begin{align}
I_{\alpha\beta\gamma\delta}  &  =\int_{\mathbf{p}}\frac{p_{\bot\alpha}%
p_{\bot\beta}p_{\bot\gamma}p_{\bot\delta}}{\Big(\bigl(p_{\bot}^{2}%
+\tau/2\bigr)^{2}+\rho^{2}p_{\Vert}^{2}\Big)^{2}}\nonumber\\
&  =\frac{3A_{\varepsilon}\mu^{-\varepsilon}}{16\rho\varepsilon}S_{\alpha
\beta\gamma\delta}+O(\varepsilon^{0})\,, \label{IntI}%
\end{align}
with $S_{\alpha\beta\gamma\delta}=(\delta_{\alpha\beta}\delta_{\gamma\delta
}+\delta_{\alpha\gamma}\delta_{\beta\delta}+\delta_{\alpha\delta}\delta
_{\beta\gamma})/3$.%
\begin{align}
J_{\alpha\beta\gamma\delta}  &  =\int_{\mathbf{p}}\frac{(p_{\Vert})^{2}%
p_{\bot\alpha}p_{\bot\beta}p_{\bot\gamma}p_{\bot\delta}}{\Big(\bigl(p_{\bot
}^{2}+\tau/2\bigr)^{2}+\rho^{2}p_{\Vert}^{2}\Big)^{3}}\nonumber\\
&  =\frac{3A_{\varepsilon}\mu^{-\varepsilon}}{64\rho^{3}\varepsilon}%
S_{\alpha\beta\gamma\delta}+O(\varepsilon^{0})\,. \label{IntJ}%
\end{align}%
\begin{align}
K_{\alpha\beta\gamma\delta}  &  =\int_{\mathbf{p}}\frac{(p_{\Vert})^{4}%
p_{\bot\alpha}p_{\bot\beta}p_{\bot\gamma}p_{\bot\delta}}{\Big(\bigl(p_{\bot
}^{2}+\tau/2\bigr)^{2}+\rho^{2}p_{\Vert}^{2}\Big)^{4}}\nonumber\\
&  =\frac{3A_{\varepsilon}\mu^{-\varepsilon}}{128\rho^{5}\varepsilon}%
S_{\alpha\beta\gamma\delta}+O(\varepsilon^{0})\,. \label{IntK}%
\end{align}

The $2$-leg diagrams are shown in Fig.~(\ref{fig:2-leg}). With the results
above we obtain their singular parts that constitute the counter-terms%
\begin{align}
\mathfrak{C}(2(a))  &  =-\frac{f}{2}\sum_{\alpha\beta\gamma\delta}q_{\bot\alpha}%
q_{\bot\beta}S_{\alpha\beta\gamma\delta}H_{\gamma\delta}\nonumber\\
&  =\frac{u}{6\varepsilon}\tau q_{\bot}^{2}\,, \label{Ct2a}%
\end{align}%
\begin{align}
\mathfrak{C}(2(b))  &  =\frac{g^{2}}{9}\sum_{\alpha\beta}q_{\bot\alpha}q_{\bot\beta
}(F_{\alpha\beta}+G_{\alpha\beta})\nonumber\\
&  =-\frac{v^{2}}{36\varepsilon}\bigl(\tau+\frac{1}{2}q_{\bot}^{2}%
\bigr)q_{\bot}^{2}\,, \label{Ct2b}%
\end{align}
and%
\begin{equation}
\mathfrak{C}(2(c))=\frac{g^{2}}{18}q_{\Vert}^{2}\sum_{\alpha\beta}I_{\alpha\alpha
\beta\beta}=\frac{v^{2}}{36\varepsilon}q_{\Vert}^{2}\,. \label{Ct2c}%
\end{equation}

Our calculation of the $3$-leg diagrams (Fig.~(\ref{fig:3-leg})) yields
\begin{align}
\mathfrak{C}(3(a))  &  =-\frac{igf}{2}q_{3,\Vert}\sum_{\alpha\beta\gamma\delta\mu
}q_{1,\bot\alpha}q_{2,\bot\beta}S_{\alpha\beta\gamma\delta}I_{\gamma\delta
\mu\mu}\nonumber\\
&  =\frac{u}{6\varepsilon}igT(\mathbf{q}_{1},\mathbf{q}_{2},\mathbf{q}_{3})\,,
\label{Ct3a}%
\end{align}
and%
\begin{align}
\mathfrak{C}(3(b))  &  =\frac{4ig^{3}}{9}q_{3,\Vert}\sum_{\alpha\beta\gamma\delta
}q_{1,\bot\alpha}q_{2,\bot\beta}J_{\alpha\beta\gamma\gamma}\nonumber\\
&  =\frac{v^{2}}{36\varepsilon}igT(\mathbf{q}_{1},\mathbf{q}_{2}%
,\mathbf{q}_{3})\,, \label{Ct3b}%
\end{align}
after symmetrization of the external leg moments. From the $4$-leg diagrams,
Fig.~(\ref{fig:4-leg}) we obtain%
\begin{align}
\mathfrak{C}(4(a))  &  =\frac{3f^{2}}{2}\sum_{\alpha\beta\gamma\delta\mu\nu\kappa
\lambda}q_{1,\bot\alpha}q_{2,\bot\beta}S_{\alpha\beta\mu\nu}I_{\mu\nu
\kappa\lambda}\nonumber\\
&  \qquad\qquad\qquad\qquad\times S_{\kappa\lambda\gamma\delta}q_{3,\bot
\gamma}q_{4,\bot\delta}\nonumber\\
&  =\frac{3u}{8\varepsilon}fS(\mathbf{q}_{1},\mathbf{q}_{2},\mathbf{q}%
_{3},\mathbf{q}_{4})\,, \label{Ct4a}%
\end{align}%
\begin{align}
\mathfrak{C}(4(b))  &  =-\frac{8fg^{2}}{3}\sum_{\alpha\beta\gamma\delta\mu\nu}%
q_{1,\bot\alpha}q_{2,\bot\beta}J_{\alpha\beta\mu\nu}\nonumber\\
&  \qquad\qquad\qquad\qquad\times S_{\mu\nu\gamma\delta}q_{3,\bot\gamma
}q_{4,\bot\delta}\\
&  =-\frac{5v^{2}}{36\varepsilon}fS(\mathbf{q}_{1},\mathbf{q}_{2}%
,\mathbf{q}_{3},\mathbf{q}_{4})\,, \label{Ct4b}%
\end{align}
and%
\begin{align}
\mathfrak{C}(4(c))  &  =\frac{16g^{4}}{27}\sum_{\alpha\beta\gamma\delta}q_{1,\bot\alpha
}q_{2,\bot\beta}K_{\alpha\beta\gamma\delta}q_{3,\bot\gamma}q_{4,\bot\delta
}\nonumber\\
&  =\frac{v^{2}}{72\varepsilon}g^{2}S(\mathbf{q}_{1},\mathbf{q}_{2}%
,\mathbf{q}_{3},\mathbf{q}_{4})\,, \label{Ct4c}%
\end{align}
again after symmetrization.

\section{Additive renormalization and RGE}
\label{app:addRen}

As pointed out in Sec.~\ref{sec:scalingAndExponents}, the Green functions $G_{0,1}(\mathbf{r}_{1})$ and $G_{0,2}(\mathbf{r}_{1}
,\mathbf{r}_{2})$  cannot be renormalized entirely by multiplicative renormalization alone. Rather, they require extra additive renormalization
which leads to the inhomogeneous RGEs ($m=1,2$)
\begin{align}
\bigl(\mu\partial_{\mu}+  &  \zeta\rho\partial_{\rho}+\kappa\tau\partial
_{\tau}+\beta_{x}\partial_{x}+\beta_{y}\partial_{y}+m\kappa\bigr)G_{0,m}%
(\{\mathbf{r\}})\nonumber\\
&  =(-\tau)^{2-m}k\frac{\mu^{-\varepsilon}}{\rho}[\delta(\mathbf{r}%
_{1}-\mathbf{r}_{2})]^{m-1}\,, \label{RGE02}%
\end{align}
where $k$ is a function of the invariant coupling constants $x$ and $y$ only.
To lowest order, the additive renormalization arises from the UV-divergence of
the diagram (4a) in Fig.~(\ref{fig:4-leg}) with removed external legs and
coupling constants. This divergence is cancelled by the additive counter-term%
\begin{equation}
K=\frac{1}{2}\sum_{\alpha,\gamma}I_{\alpha\alpha\gamma\gamma}=\frac
{\mu^{-\varepsilon}}{16\pi\rho}\frac{1}{\varepsilon}+O(\varepsilon^{0})\,,
\end{equation}
which leads after applying the operator $\left.  \mu\partial_{\mu}{}_{\cdots
}\right\vert _{0}$ to $k=1/16\pi$ at lowest order.

To proceed towards its solution, let's discus the form of the RGE (\ref{RGE02}) in some more detail. This RGE is a partial differential equation
that describes the dependence of $G_{0,m}=:u_{m}(\xi)$ upon the variables $(\mu,\rho ,\tau,x,y)=:\xi=(x_{\nu};\,\nu=1,\ldots,5)$. As far as the RGE is concerned, the spatial coordinate plays solely the role of a constant parameter. The general form of
the RGE is a quasi-linear partial differential equation (PDE), see
\textit{e.g.} \cite{Kam56}%
\begin{equation}
\sum_{\nu}a_{\nu}(\xi,u_{m})\frac{\partial u_{m}}{\partial x_{\nu}}%
=b(\xi,u_{m})\,. \label{PDE}%
\end{equation}
To find its solution, we consider $u_{m}(\xi)$ as the solution of the implicit
equation
\begin{equation}
w(u_{m},\xi)=0\,,
\end{equation}
and write the PDE as%
\begin{equation}
\sum_{\nu}a_{\nu}(\xi,u_{m})\frac{\partial w}{\partial x_{\nu}}+b(\xi
,u_{m})\frac{\partial w}{\partial u_{m}}=0\,.
\end{equation}
This homogeneous PDE can be solved by employing the characteristics $\bar{\xi}(\ell)$,
$\bar{u}_{m}(\ell)$ determined as usual by the ordinary differential equations%
\begin{equation}
\ell\frac{d\bar{\xi}}{dl}=a(\bar{\xi},\bar{u}_{m})\,,\qquad\ell\frac{d\bar
{u}_{m}}{dl}=b(\bar{\xi},\bar{u}_{m})\,, \label{Char}%
\end{equation}
with $a=(a_{\nu})=(\mu,\zeta\rho,\kappa\tau,\beta_{x},\beta_{y})$, and initial
conditions $\bar{\xi}(1)=\xi$, $\bar{u}_{m}(1)=u_{m}$. Since $w$ remaines
constant along the characteristic flow, the solution is%
\begin{equation}
w(\bar{u}_{m}(\ell),\bar{\xi}(\ell))=w(u_{m},\xi)=0\,,
\end{equation}
and $\bar{u}_{m}(\ell)$ is of the same functional form as that of $u_{m}$ as a function of $\xi$,
\begin{equation}
\bar{u}_{m}(\ell)=u_{m}(\bar{\xi}(\ell))\,. \label{SolPDE}%
\end{equation}
In our case, the characteristics $\bar{\mu}(\ell)=\ell\mu$, $\bar{\rho}%
(\ell)=X_{\rho}(\ell)\rho$, $\bar{\tau}(\ell)=X_{\tau}(\ell)\tau$, $\bar
{x}(\ell)$, $\bar{y}(\ell)$ are defined by the differential equations
(\ref{floweq}) and (\ref{ampldiff}), and their asymptotic solutions in the case
$\varepsilon=0$, $\ell\rightarrow0$ are given in equations (\ref{asympt}) and
(\ref{AmplLog}). It remains to solve the differential equation for $\bar{u}_{m}(\ell)$%
\begin{align}
\ell\frac{d\bar{u}_{m}(\ell)}{dl}=\bigl(-X_{\tau}(\ell)  &  \tau
\bigr)^{2-m}\frac{k(\bar{x}(\ell),\bar{y}(\ell))}{X_{\rho}(\ell)\rho}%
\delta(\mathbf{r}_{1}-\mathbf{r}_{2})\nonumber\\
&  -m\kappa(\bar{x}(\ell),\bar{y}(\ell))\bar{u}_{m}(\ell)\,. \label{Diff-u}%
\end{align}
This equation has the solution
\begin{equation}
\bar{u}_{m}(\ell)=X_{\tau}(\ell)^{-m}\Big(u_{m}-\frac{(-\tau)^{2-m}}{\rho
}I(\ell)\delta(\mathbf{r}_{1}-\mathbf{r}_{2})\Big)\,, \label{u-l}%
\end{equation}
where $I(\ell)$ is the integral%
\begin{equation}
I(\ell)=\int_{\ell}^{1}\frac{d\ell^{\prime}}{\ell^{\prime}}\frac{X_{\tau}%
(\ell^{\prime})^{2}}{X_{\rho}(\ell^{\prime})}k(\bar{x}(\ell^{\prime}),\bar
{y}(\ell^{\prime}))\,. \label{I-l}%
\end{equation}
Since $u_{m}=G_{0,m}(\mathbf{r}_{1},\mathbf{r}_{2};\xi)$ and $\bar{u}_{m}%
(\ell)=G_{0,m}(\mathbf{r}_{1},\mathbf{r}_{2};\xi(\ell))$, we finally get 
\begin{align}
G_{0,m}(\{\mathbf{r\}};\xi)=  &  X_{\tau}(\ell)^{m}G_{0,m}(\{\mathbf{r\}}%
;\xi(\ell))\nonumber\\
&  +\frac{(-\tau)^{2-m}}{\rho}I(\ell)[\delta(\mathbf{r}_{1}-\mathbf{r}%
_{2})]^{m-1} \label{G-fin}%
\end{align}
from Eq.~(\ref{u-l}).


\begin{thebibliography}{99}
\bibitem{footnote1}Note that the so-called steering effect by which particles are deflected parallel to the surface when they reach a certain height above it leads to the same relation.
\end{thebibliography}
\end{document}